# COMMUNICATION

# A generalized kinetic framework applied to whole-cell catalysis in biofilm flow reactors clarifies performance enhancements

Mir Pouyan Zarabadi,[a] Manon Couture,[b, d] Steve J. Charette,[b, c, d] Jesse Greener*,[a, e]

**Abstract:** A common kinetic framework for studies of whole-cell catalysis is vital for understanding and optimizing bioflow reactors. In this work, we demonstrate the applicability of a flow-adapted version of Michaelis-Menten kinetics to a catalytic bacterial biofilm. A three-electrode microfluidic electrochemical flow cell measured increased turnover rates by as much as 50% from a *Geobacter sulfurreducens* biofilm as flow rate was varied. Based on parameters from the applied kinetic framework, flow-induced increases to turnover rate, catalytic efficiency and device reaction capacity could be linked to an increase in catalytic biomass. This study demonstrates that a standardized kinetic framework is critical for quantitative measurements of new living catalytic systems in flow cells and for benchmarking against well-studied catalytic systems such as enzymes.

For centuries, whole-cell biotransformations have been used in the production of food and beverages. More recently, whole-cell biocatalysis has become an active area of research[1] due to its potential for low-cost synthesis of fine chemicals, including chiral molecules and pharmaceuticals,[2] natural food additives,[3] and applications in bioremediation and energy.[4] Whole-cell biocatalysis benefits from complex multi-enzyme reaction steps, applicability at ambient conditions, and attractive properties such as self-repair and regeneration. It can also be lower in cost compared with extracted enzymes because there is no need for isolation and purification steps and because cells produce their own enzyme co-factors.[5] Bacterial biofilm are promising for the same reasons, but include other benefits than make them candidates for industrial processes as well.[6] These include their preference for surface attachment, making them ideal for heterogeneous catalysis and because of their protective self-produced extracellular polymeric matrix, which can mitigate challenges related to toxicity.

Control over substrate concentration and reactor feeding strategy are the most important factors for optimisation of whole-cell biocatalysis.[7,8] Chemostat bioreactors can impose tunable concentrations against the biofilm and eliminate cyclic nutrient depletion and product accumulation between solution replenishment in bulk reactors.[9] However, latency in manipulations of reaction conditions during reaction optimization and fundamental research is a drawback. Microfluidic channels can address this problem because dead is space nearly zero. Combined with strictly laminar flow, microfluidic bioreactors also offer precise control of the shear forces and of diffusion barriers at the biofilm-liquid interface.[10] High surface-area-to-volume ratios result in efficient diffusive mass-transfer between the nutrient solution and the wall-adhered biofilm, even for short contact times at high flow velocities.[11] Due to the small volumes used in microfluidic bioreactors, liquid consumption is reduced and thus long-duration studies are possible under a large range of hydrodynamic conditions without the need for frequent refilling of reagent sources. With the proliferation of microfluidic bioanalytical tools for real-time *in situ* measurements of biofilms and their by-products, deeper scientific investigations are possible with better accuracy and repeatability.[10] In this work, we use microfluidic electrochemical flow cells to study whole-cell biofilm electrocatalysis under flow. The kinetic framework used connects whole-cell catalysis in flow systems to traditional Michaelis-Menten kinetics (Eq. 1) between a substrate molecule S and a catalytic bacteria in a biofilm, denoted by E, due to historical relation to enzyme kinetics.[12,13]

$$E + S \underset{k'_a}{\overset{k_a}{\rightleftharpoons}} ES \overset{k_{cat}}{\rightarrow} P + E \qquad (Eq.\ 1)$$

where $k_a$ (M$^{-1}\cdot$s$^{-1}$) and $k'_a$ (s$^{-1}$) are the forward and backward rate constants in the first reversible complexation step, and $k_{cat}$ is the catalytic rate constant (s$^{-1}$) that irreversibly transforms the pre-equilibrated ES complex into the products. The irreversibility of the final transformation is a typical assumption for most biocatalytic systems but should be justified. The Michaelis-Menten constant $K_M$ defines the substrate concentration that results in half-maximal activity. The reader is directed to the supporting information (SI, section 4) for more information.

Changes to certain reaction conditions can result in "apparent" or "effective" kinetics, which lead to the apparent Michaelis-Menten constants and efficiency parameters $K_{M(app)}$ and $\varepsilon_{(app)}$, respectively. In 1966, Lilly and Hornby developed a framework for enzyme kinetics that accounted for changes to hydrodynamic conditions in flow reactors[14] (Eq. 2).

$$P[S]_i = \frac{C}{Q} + K_{M(app)} \ln(1 - P) \qquad (Eq.\ 2)$$

where $P \left(= \frac{[S]_i - [S]_f}{[S]_i}\right)$ is the substrate conversion fraction, and $C$ ( = $k_{cat}[E]\beta$) is known as the device reaction capacity, which includes ratio of the reactor void volume to the total reactor volume, $\beta$. Assuming that $k_{cat}$ and $\beta$ are not sensitive to flow, changing C can supply evidence of forced convection through three-dimensional porous systems such as biofilms, a point that is often overlooked for reactive biofilms. Applied to enzyme catalysis, trends in $K_{M(app)}$ are often extrapolated to zero flow $K^0_{M(app)}$ to determine whether enzyme activity is only affected by mass transfer effects $K_M$ = 

[a] Mirpouyan Zarabadi, Prof. Jesse Greener
Département de Chimie, Faculté des sciences et de génie, Université Laval, Québec City, QC, Canada.
E-mail: jesse.greener@chm.ulaval.ca
[b, d] Prof. Steve J. Charette, Prof. Manon Couture
Institut de Biologie Intégrative et des Systèmes, Université Laval, Québec City, QC, Canada.
Département de biochimie, de microbiologie et de bio-informatique, Faculté des sciences et de génie, Université Laval, Québec City, QC, Canada.
[c] Prof. Steve J. Charette
Centre de recherche de l'Institut universitaire de cardiologie et de pneumologie de Québec, Québec City, QC, Canada.
[e] Prof. Jesse Greener
CHU de Québec, centre de recherche, Université Laval, 10 rue de l'Espinay, Québec, QC, Canada.

Supporting information for this article is given via a link at the end of the document.((Please delete this text if not appropriate))

$K^0_{M(app)}$ or whether surface immobilization can also change the kinetics.[15-17] For enzymes in their native environment within whole cells or biofilms, it is expected that the $K_M$ value measured during static experiments should be the same as $K^0_{M(app)}$ from flow reactors.

Although Michaelis-Menten kinetics have been applied to whole-cell catalysis, the Lilly-Hornby approach has not, limiting the advancement of whole-cell catalysis in flow reactors. To address this deficiency, the turnover rate ($\frac{d[P]}{dt} = -\frac{d[S]}{dt}$, Eq. S2 in SI) from surface-adhered cells or their biofilms should be measured while accurate control over flow rate and [S] are applied. An electroactive biofilm (EAB) from electrogenic bacteria, such as *Geobacter sulfurreducens* is an excellent choice because it can transfer electrons to an electrode during respiration,[10] enabling direct observations of instantaneous turnover rate using chronoamperometric measurements of electric current. For a monoculture EAB of *G. sulfurreducens* under anaerobic conditions, the turnover rate of an acetate substrate (Ac), namely $\frac{dmol_{Ac}}{dt}$ (mol$_{Ac}$·s$^{-1}$), is calculated from the electrical current I (C·s$^{-1}$) using Eq. 3:

$$\frac{dmol_{Ac}}{dt} = \frac{I}{8 \cdot F} \quad \text{(Eq. 3)}$$

where F is Faraday's constant or F = 9.6485 × 10$^4$ C·mol$_e^{-1}$, and 8 is the proportionality constant for the number of moles of electrons produced for each mole of Ac oxidized. The irreversible final transformation in Eq. 1 is a good assumption for chronoamperometry experiments on electrode-adhered EABs due to the application of an electrode potential, which can quickly and efficiently conduct electrons away from the reaction site.

In this work, we conducted real-time measurements of the instantaneous turnover rates from EAB of *G. sulfurreducens* bacteria in a microfluidic three-electrode flow cell under controlled reaction conditions. Figure 1a presents the top- and side-view schematics of the device, including the sequential placement of the electrodes. Figure 1b shows a typical scanning electron microscopy image of a mature biofilm on the WE following an experiment. The reader is directed to the SI for additional details on device fabrication, operating conditions, and preparation of *G. sulfurreducens*. The reader is also directed to the SI (Section 5) for a justification that electron transfer kinetics are not rate limiting under different flow conditions. Thus turnover rate can be interpreted substrate conversion via a modified version of Michaelis-Menten kinetics.[18]

$$I = I_{max} \frac{[Ac]}{K_{M(app)} + [Ac]} \quad \text{(Eq. 4)}$$

where $I_{max}$ is the maximum current output when the substrate [Ac] is much larger than $K_{M(app)}$.

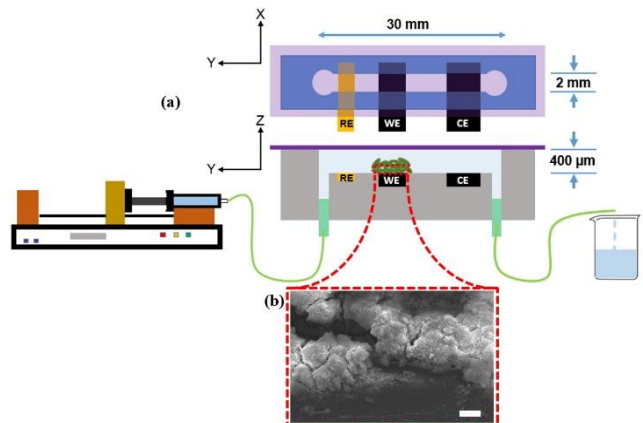

**Figure 1 (a)** Schematic of the three-electrode electrochemical microfluidic flow channel with channel dimensions 2 mm (w) × 0.4 mm (h) × 30 mm (L). The device was fabricated in PDMS with a glass sealing layer (purple). The top view (x-y plane) through the glass layer (purple) and the side view (y-z plane) show the PDMS channel (grey, or blue as viewed through the glass), the embedded gold reference electrode (RE), and the graphite counter and working electrodes (CE and WE, respectively). The side view shows fluidic connections to the 10 mL gas-tight syringe and pump and the waste and growth of the G. sulfurreducens (green) at the WE surface. **(b)** Scanning electron microscope image of mature G. sulfurreducens biofilm (scale bar: 20 µm).

The chronoamperometry data in Figure 2 show peaks in the current I over a background signal of approximately 20 µA as flow pulses ranging from Q = 0.4 to 3 mL·h$^{-1}$ were applied over a background flow rate of Q = 0.2 mL·h$^{-1}$. The Figure 2 inset plot presents the averaged I vs. Q results from four different measurements on different days. Flow rate of Q = 3 mL·h$^{-1}$ resulted in a 24% increase in current measurements over those acquired at Q = 0.4 mL·h$^{-1}$. Enhancements could not be compared with direct measurements under static conditions (Q = 0) because continuous Ac depletion in small microchannel volume around the electrode would prevent stable measurements of current. However, an extrapolation to Q = 0 using current values in the linear region from Q = 0.2 to Q = 1 mL·h$^{-1}$ gave an estimated zero flow current value of I = 18.25 µA (3 A·m$^{-2}$), which is in the normal range (1-10 A·m$^{-2}$) for other reported experiments under static conditions. Therefore, we estimated that the turnover rate at 3 mL·h$^{-1}$ was more than 50% higher than that under static conditions. Based on the electric current obtained for each flow rate Q, the [Ac] conversion was calculated for different initial acetate concentrations [Ac]$_i$ in the range of 0.3 to 10 mM using the formula in Eq. 5:

$$P \cdot [Ac]_i = [Ac]_i - [Ac]_f = \frac{dmol_{Ac}}{dt} / Q \quad \text{(Eq. 5)}$$

where P is defined for Eq. 2, and [Ac]$_f$ is the final Ac concentration after biocatalytic oxidation. Figure 3a shows separate plots of the change in [Ac] as a function of Q for the different applied [Ac]$_i$. According to Eq. 2, as [Ac]$_i$ is varied, the plot of P·[Ac]$_i$ vs. – Ln (1-P) should result in a straight line with slope - $K_{M(app)}$ and intercept C/Q. Such a linear plot was obtained for the five different flow rates used in this work (Figure 3b).

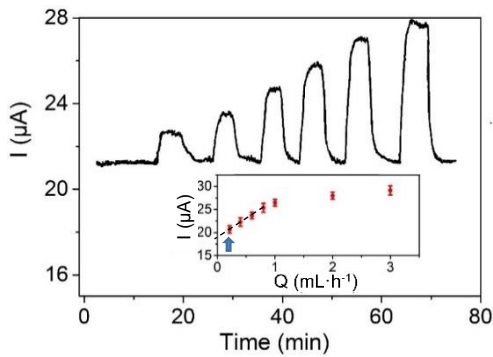

**Figure 2** Flow rate modulation from Q = 0.2 mL·h$^{-1}$ (base flow) to elevated values, Q = 0.4, 0.6, 0.8, 1, 2 and 3 mL·h$^{-1}$ for a 600 h old *G. sulfurreducens* biofilm exposed to [Ac] = 10 mM. Inset: average I vs. Q during for a mature biofilm (> 600 h) conducted 4 times on four consecutive days. Error bars represent standard deviation of averaged measurements. The blue arrow points to data at Q = 0.2 mL/h that were acquired from background current measurements, whereas the other values were acquired from peak current values in the main figure. The dashed line extrapolates the linear portion of the inset figure to Q = 0 conditions.

Figure 3c shows the values of $K_{M(app)}$ obtained from the slopes in Figure 3b. The trends in $K_{M(app)}$ are discussed after validation of the technique by comparison to the conventional $K_M$ constant from static experiments. First, we extrapolated to Q = 0, obtaining $K^0_{M(app)}$ = 0.59 mM. Then, direct measurements of $K_M$ were collected from a *G. sulfurreducens* biofilm in a bulk three-electrode system under similar conditions (bacterial age, graphite electrode material and nutrient solution). Electric currents were obtained immediately after stabilization following replacement with a new nutrient solution with a different [Ac] (SI, Figure S5b). Plotting I vs. [Ac] produced a standard Michaelis-Menten profile (Figure 3d) and a Lineweaver-Burk plot of 1/I vs. 1/[Ac] yielded the expected straight line (Figure 3d inset). A fitting algorithm applied to either curve in Figure 3d yielded a value of $K_M$ = 0.62 mM. The similarity between $K_M$ and $K^0_{M(app)}$ for the present system also matched the $K_M$ = 0.60 mM reported previously for bulk measurements.[18]

The implications of reduced $K_{M(app)}$ with increasing Q are examined next. Applied Ac concentrations of 10 mM are considered to lie in the "substrate saturated" regime[19] ([Ac] >> $K_M$), and should be described by Eq. S3 (equivalently I=$k_{cat}$[E]). Therefore, an increase in either or both of $k_{cat}$ or [E] could explain the peak I values (Figure 2), which increased by 24% when Q was increased from 0.4 to 3 mL·h$^{-1}$. Despite previous suggestions that accelerated turnover under flow could be the result of EAB deacidification,[20] a likely route to increases in $k_{cat}$, pH was recently shown not to change for similar conditions at [Ac] = 10 mM.[20] Alternatively, flow-induced increases to current might be at least partially related to changes to [E]. In this vein, consider the measurements of device reaction capacity (C), which increased by 19% when increasing flow from Q = 0.4 to Q = 3 mL·h$^{-1}$ (SI, Table S1 and Figure S6). Taking into account that C depends only on [E], $k_{cat}$, and the physical dimensions implicit in β (Eq. 2), flow-related increases to C support the likelihood of increases to [E], assuming that $k_{cat}$ and β remain largely constant. Flow-based increases to [E] could result from better contact between the acetate molecules and the catalytic bacteria at different strata within the biofilm due to forced convective flow through the porous biofilm, as noted previously for non-electroactive biofilms.[22, 23] The implication of additional *G. sulfurreducens* bacteria contributing to biocatalysis combined with reduced diffusion barriers around the biofilm at higher flow rates, can be exploited for improvement to device performance in future flow-based bioelectrochemical applications. Global improvements in performance under flow are expressed by increases to (apparent) catalytic efficiency $\varepsilon_{(app)}$ = $k_{cat}/K_{M(app)}$, by replacing $K_M$ with $K_{M(app)}$ in Eq. S4. Notably, increases to $\varepsilon_{(app)}$ were calculated to be 19% as Q increased from 0.4 to 3 mL·h$^{-1}$, which correspond closely to measured changes to I and C and proposed changes in [E] for the same flow rate range.

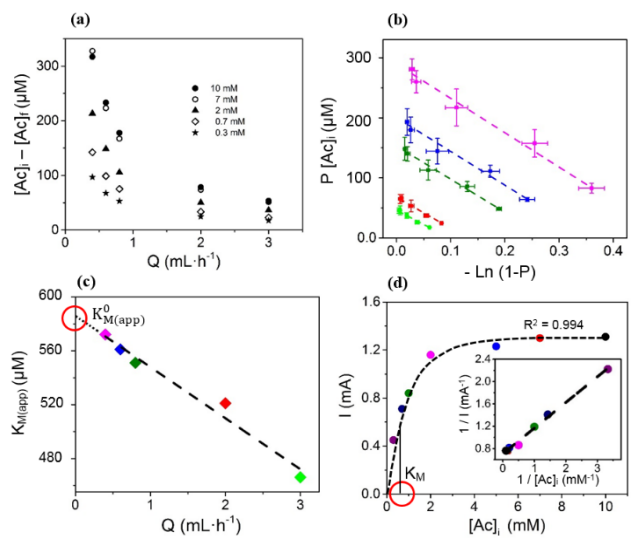

**Figure 3 (a)** Plots of the concentration of acetate nutrient converted for respiration vs. flow rate as a function of the initial acetate concentration. **(b)** Plots of P [Ac]$_i$ vs. – Ln(1 - P). The modulated flow rates were Q = 0.4 mL·h$^{-1}$ (pink), 0.6 mL·h$^{-1}$ (blue), 0.8 mL·h$^{-1}$ (dark green), 2 mL·h$^{-1}$ (red) and 3 mL·h$^{-1}$ (green). **(c)** Plot of $K_{M(app)}$ vs. Q. The intercept on the vertical axis (red circle) yields a zero-flow $K_{M(app)}$ value ($K^0_{M(app)}$) of 0.59 mM. **(d)** Current vs. initial acetate concentrations in the bulk experiment for 10 mM (black), 7 mM (red), 5 mM (blue), 2 mM (pink), 1 mM (dark green), 0.7 mM (dark blue) and 0.3 mM (purple). The curve was fitted to Eq. 4 to find $K_{M(app)}$. Inset: The Lineweaver-Burk plot of the reciprocal current output vs. the reciprocal acetate concentration demonstrates the expected linear profile for systems applicable to Michaelis-Menten kinetics.

As applied to the study of *G. sulfurreducens* biofilms, a flow-adapted version of the Michaelis-Menten equation was used for the first time to understand and develop whole-cell catalysis while leveraging the advantages of microfluidic flow cells. Various applications can be envisioned, especially those related to synthetic biology and metabolic engineering in which cells are effectively considered as units of production for valuable products of bulk and fine chemicals.[24] In addition, characterization of enzymes within their native (cellular) environment is an important goal in the field of enzymology.[25] Our work shows that measurements of kinetic parameters such as $K_M$, $\varepsilon_{(app)}$, and [E] as a function of flow for whole-cell catalysts is feasible microorganisms immobilized within a microfluidic chamber. Generalization of this approach together with the application of different detection modes for non-electroactive biofilms appears a desirable next step.

## Experimental Section

Frozen samples of *Geobacter sulfurreducens* (strain PCA, ATCC 51573), were cultured under controlled temperature and deoxygenated conditions before being injected into the microfluidic electrochemical device. The channel-embedded electrodes were fabricated as shown previously.[20,26] Graphite was used for both the working (WE) and counter (CE) electrodes and a stable gold (Au) pseudo reference electrode (RE). The device was operated in an anaerobic enclosure with controlled temperature of 23 ± 0.5 °C. Syringe pumps were used for controlled liquid injection while electrochemical measurements were conducted using a potentionstat for chronoamperometry and cyclic voltammetry.

## Acknowledgements

This research was supported by funding from the Natural Sciences and Engineering Research Council, Canada and Sentinelle Nord. J.G. is the recipient of an Early Researcher Award and an AUDACE grant (high risk, high reward) for studies of microbiological systems using microfluidics from the Fonds de recherche du Québec—nature et technologies (FRQNT). The authors wish to thank Molly Gregas for copyedits.

**Keywords:** whole-cell catalysis • Lilly-Hornby • biofilm flow reactor • electrochemistry • microfluidics

Supporting information

# A generalized kinetic framework for whole-cell catalysis in flow reactors clarifies performance enhancements

*Mir Pouyan Zarabadi[1], Manon Couture[2,4], Steve J. Charette[2,3,4] and Jesse Greener[\*,1,5]*

[1] Département de Chimie, Faculté des sciences et de génie, Université Laval, Québec City, QC, Canada.

[2] Institut de Biologie Intégrative et des Systèmes, Université Laval, Québec City, QC, Canada.

[3] Centre de recherche de l'Institut universitaire de cardiologie et de pneumologie de Québec, Québec City, QC, Canada.

[4] Département de biochimie, de microbiologie et de bio-informatique, Faculté des sciences et de génie, Université Laval, Québec City, QC, Canada.

[5] CHU de Québec, centre de recherche, Université Laval, 10 rue de l'Espinay, Québec, QC, Canada.

\* E-mail : jesse.greener@chm.ulaval.ca

**Sections :**

1. Bacterial preparation
2. Device fabrication, anaerobic environment and inoculation
3. SEM sample preparation
4. The Michaelis-Menten and Lilly-Hornby kinetics
5. Verification of efficient electron transfer kinetics
6. Initial biofilm growth in microfluidic electrochemical cell
7. Bulk set-up for bacterial growth and measurements of $K_M$
8. Influence of flow on reactor parameters
9. Trends in the device reaction capacity
10. Trends in $K_{M(app)}$ and reaction mechanisms
11. References

1. **Bacterial preparation:**

Frozen samples of *Geobacter sulfurreducens* (strain PCA, ATCC 51573) were cultured at 30 °C using an anaerobic medium in an anaerobic chamber with 10% $H_2$ and 10% $CO_2$ balanced with $N_2$ for 7 days and sub-cultured at least 2 times prior to injection into the electrochemical device. The medium contained the following (per liter of distilled water): 1.5 g $NH_4Cl$, 0.6 g $NaH_2PO_4$, 0.1 g KCl, 2.5 g $NaHCO_3$, 0.82 g $CH_3COONa$ (acetate, 10 mM), 8 g $Na_2C_4H_2O_4$ (fumarate, 40 mM), 10 mL vitamin supplements ATCC® MD-VS™, 10 mL trace mineral supplements ATCC® MD-TMS™. With the exception of sodium fumarate and vitamin/trace mineral supplements that were added after filter sterilization into final nutrient medium, all chemical compounds were dissolved in distilled water and sterilize by autoclaving at 110 °C and 20 psi for 20 min. The nutrient medium was adjusted to pH = 7 and *G. sulfurreducens* were sub-cultured 3 to 8 times. Sodium fumarate and vitamins were added in the nutrient solution only for biofilm growth of planktonic bacteria, whereas they were removed to encourage electrode respiration.

2. **Microfluidic device fabrication, anaerobic environment and inoculation:**

The microfluidic device fabrication, anaerobic condition, inoculation and bacterial preparation have been explained in previous publications,[1, 2] and are briefly reviewed here with emphasis on the present context.

**Device fabrication (electrodes):** The microfluidic electrochemical cell used in this study included a three-electrode configuration, consisting of graphite working (WE) and counter (CE) electrodes and a gold pseudo reference electrode (RE). The WE and CE were cut from a commercial source (GraphiteStore.com Inc., USA) into 3×20 and 4×20 mm strips, respectively. The pseudo reference electrode (RE) was created by Au electroless deposition on a polystyrene support material and was subsequently cut into 3×20 mm strips.

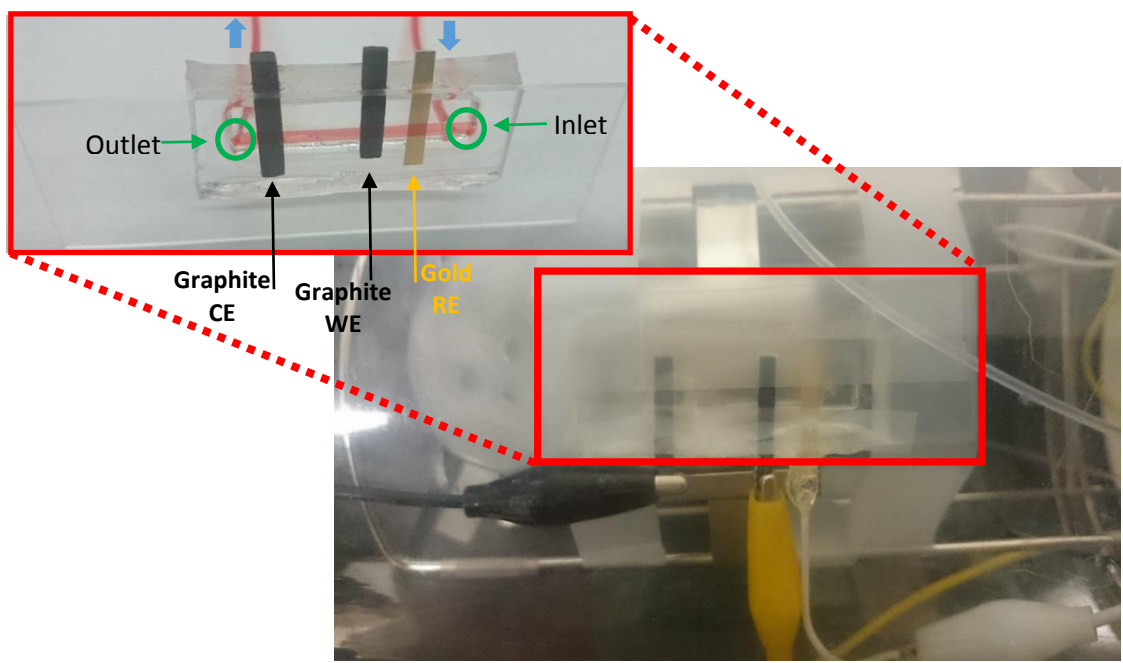

**Figure S1.** Image of the three-electrode device after installation within the anaerobic enclosure. Electrical connections to the counter electrode and working electrode are shown via black and yellow alligator clips. The epoxy-protected solder connection to the gold reference electrode is shown with the white wire. Inset: close up of the microfluidic device before installation and electrical connections with the red dye solution being passed through the channel for contrast.

**Device fabrication (electrodes integration):** a typical silicon master mould with SU8 features (FlowJEM Inc., Toronto, Canada) defined the channel length, width, height dimensions being 30 (L), 2 (w), 0.4 mm (h). Electrodes were wrapped entirely with tape to prevent PDMS from directly contacting any portion of the electrodes. Then, electrodes were embedded into the microchannel by first placing them flush against the top of the channel feature on the silicon master mould and held in place with double-sided tape. A mixture of liquid polydimethylsiloxane (PDMS) and cross-linking agent Sylgard184 (Dow Corning, Canada) (ratio 10:1) was poured over the mould with the electrodes in place. After curing the PDMS for 4 h at 70 °C, the electrodes were embedded at the bottom of the channel and the device was removed from the mould. The residual tape was cut from all the electrode surfaces. Before sealing the channel, graphite electrodes were swabbed 2 M

HCl to remove any debris following the channel fabrication. Then the entire channel was cleaned and sterilized using a 70% ethanol solution and autoclaved distilled water. The microfluidic electrochemical cell was then sealed with a microscope slide by exposure to air plasma (PCD-001 Harrick Plasma, Ithaca, NY, USA). Electrical connections were made to the exposed parts of electrodes outside of the device. The gold RE was connected by solder and a protective coating of epoxy was added to physically stabilize the connection. The electrical connections to the graphite electrodes were accomplished with alligator clips. The finalized microfluidic device is shown in Figure S1 (inset).

**Anaerobic enclosure:** The PDMS polymer enables diffusion of small molecules including $O_2$ through it. Therefore, the microfluidic device was housed within a small anaerobic enclosure (McIntosh and Filde's, 28029 Sigma-Aldrich) and filled with an anaerobic gas (20% $CO_2$ and 80% $N_2$). A feedthrough port in the enclosure enabled electrical connections between electrodes and potentiostat and the sterile perfluoroalkoxy connective tubing (PFA tube 1/16, Hamilton Inc., Canada) between the device liquid connectors and syringe pumps. The port was sealed using epoxy glue to prevent air exchange with the ambient conditions. A layer of epoxy glue with gas-impermeable tape was applied on the tubing to minimize gas diffusion through the connective tubing outside of the enclosure. The inlet tube was connected to a 50 mL glass syringe and 10 mL for inoculations via connector assemblies (P-200x, P-658, IDEX, WA, USA). Syringe pumps (PHD 2000, Harvard Apparatus, Holliston, MA, USA) were used in controlled liquid injection. Temperature control of 23 ± 0.5 °C was verified by a local temperature probe. A picture of the microfluidic device installed in the anaerobic chamber, is shown in Figure S1.

**Inoculation:** Before inoculation, the microfluidic channel and tubing were rinsed with sterile distilled water for 1 h at 1 mL·h$^{-1}$. All inoculation and subsequent biofilm growth was conducted in a, in a 20:80 $CO_2$:$N_2$ gas purged system ensuring constant anaerobic conditions. A 1.5 mL inoculum solution containing *G. sulfurreducens* was first introduced to the microchannel through for 3 h at Q = 0.5 mL·h$^{-1}$. The medium solution contained dissolved fumarate to enable extracellular electron transfer by planktonic bacteria. During inoculation, an oxidative electrochemical potential was applied to the working electrode (400 mV vs. Au, equivalent to 0 mV vs. Ag/AgCl)[2] which enabled electrode respiration for sessile bacteria. After 3 h inoculation process, the inoculum syringe was exchanged for a 50 mL air-tight syringe containing acetate nutrient medium with no sodium fumarate or vitamins, with flow rate Q = 0.2 mL·h$^{-1}$. Exclusion of fumarate (and to a less important degree, vitamins) ensured that external electron transfer could only occur at the WE, which remained poised at 400 mV vs. Au. The system was maintained under these growth conditions, with periodic replacement of the nutrient syringe for the duration of the experiment. The current output was monitored by chronoamperometry during growth of *G. sulfurreducens* electroactive biofilm.

## 3. SEM sample preparation

Scanning electron microscopy (SEM) was conducted after the experiment finished with the objective to observe the bacteria attached to the WE (Figure 1b). Before the electrode was removed from the microfluidic device, a fixation solution (2.5 % glutaraldehyde in phosphate buffer) was applied to channel (Q = 0.5 mL·h$^{-1}$) for 2 h while the device was still under anaerobic conditions. Then, the device was removed from the anaerobic enclosure and cut open to expose and remove the WE. The biofilm coated electrode was then left exposed to the same fixation solution in a bath overnight. The next day the electrode was then transferred to a solution with 1 % osmium tetroxide for 1.5 h and rinsed in phosphate buffer. Finally, the sample was sequentially dehydrated in 50, 75, 95 and 100% aqueous ethanol solutions for 15 min, respectively, followed by room temperature drying overnight. Before acquiring images, a thin layer of gold was sputtered on the biofilms and electrode (Model: Nanotech, SEM PREP 2). The images were captured with JEOL JSM-6360 LV electron microscope imaging platform.

## 4. The Michaelis-Menten and Lilly-Hornby kinetics

This section contains information relating the classic Michaelis-Menten equation and its flow variant, Lilly-Hornby. A key parameter in the Michaelis-Menten equation is the Michaelis-Menten constant $K_M$, otherwise known as the half-saturation concentration. $K_M$ is calculated in based on the individual rate constants from Eq. 1 based on Eq. 1S:

$$K_M = \frac{k'_a + k_{cat}}{k_a} \quad \text{(Eq. S1)}$$

In the application of this model to whole-cell kinetics, each reaction step is complex and consists of several sequential biochemical enzymatic reactions, e.g., in the tricarboxylic acid cycle or TCA cycle, as well as molecular diffusion steps. Nevertheless, the Michaelis-Menten model gives a useful framework for understanding overall kinetics, which can be described by the rate equation in Eq. S2:

$$\frac{d[P]}{dt} = \frac{d[P]}{dt}\bigg|_{max} \frac{[S]}{K_M + [S]} \quad \text{(Eq. S2)}$$

where $\frac{d[P]}{dt}$ (mol·L$^{-1}$·s$^{-1}$) is the rate of product production or equivalently, the turnover rate. When [S] = $K_M$, the reaction proceeds at half-maximal rate ($\frac{1}{2}\frac{d[P]}{dt}\big|_{max}$) and tends to a maximum ($\frac{d[P]}{dt}\big|_{max}$), often referred to as the maximum velocity, when [S] >> $K_M$. Under these conditions, the maximum rate is related to $k_{cat}$ and [E] via Eq. S3:

$$\frac{d[P]}{dt}\bigg|_{max} = k_{cat}[E] \quad \text{(Eq. S3)}$$

Finally, the reaction efficiency $\varepsilon$ can be estimated from Eq. S4:

$$\varepsilon = \frac{k_{cat}}{K_M} \quad \text{(Eq. S4)}$$

Lilly and Hornby obtained Eq. 2 (main paper) by integrating the Michaelis-Menten rate equation with the boundary conditions for substrate concentrations at the inlet [S]$_i$ and outlet [S]$_f$, thereby yielding Eq. S5:

$$([S]_i - [S]_f) - K_{M(app)} \ln \frac{[S]_f}{[S]_i} = k_{cat} \left(\frac{[E]}{Q}\right)(\beta) \quad \text{(Eq. S5)}$$

where $\beta$ is the ratio of the reactor void volume to the total reactor volume. With knowledge of the reactor conditions, Eq. S5 can be written in the form shown in Eq. S6 to describe $K_{M(app)}$ as a function of the flow rate, Q (mL·s$^{-1}$).[4]

$$([Ac]_i - [Ac]_f) - K_{M(app)} \ln \frac{[Ac]_f}{[Ac]_i} = k_{cat} \left(\frac{[E]}{Q}\right) (\beta) \qquad \text{(Eq. S6)}$$

Equation S6 is equivilant to the Eq. 2 in the main paper based on the definition of substrate conversion fraction, $P = \frac{[S]_i - [S]_f}{[S]_i}$, and the device reaction capacity, $C = k_{cat} [E] \beta$.

## 5. Verification of efficient electron transfer kinetics

To attribute the changes in electric current to Michaelis-Menten kinetics during manipulations of flow, the electron transfer rate must not be rate-limiting compared with the nutrient conversion step. In addition to the use of graphite WE and CE to improve charge transfer and efficient oxygen reduction reactions over long experimental durations,[7] a direct evaluation of the role of electron transfer kinetics was undertaken. Blending the The Nernst and Michaelis–Menten equations has been used to model the steady-state kinetics of an enzyme/electrode redox systems, including for electroactive biofilms[8]. For such a system where $[Ac] \gg K_M$, Michaelis-Menten kinetics result in a maximum current, $I_{max}$, yielding Eq. S7, which shows the role of applied potential (E) on the current.

$$I = I_{max} \left\{ \frac{1}{1 + \exp\left[-\frac{F}{RT}(E - E_{KA})\right]} \right\} \qquad \text{(Eq. S7)}$$

where R is the ideal gas constant (8.3145 J.mol$^{-1}$.K$^{-1}$), F is the Faraday constant (9.6485 x10$^4$ C.mol$^{-1}$), T is the temperature (K), and $E_{KA}$ (V) is related to the standard reduction potential of proteins associated with respiration. Applying a potential $E=E_{KA}$ gives the half maximum activity, resulting in $I = 1/2 \cdot I_{max}$ somewhat similar to the role of $K_M$ for concentration in the Michaelis-Menten kinetics. Under ideal behaviour, current should follow the typical S-curve generated by the Nernst term in Eq. S1. Moreover, if $E \gg E_{KA}$, the electron transfer is rapid and not limiting, causing the term to reduce to 1. In this case the metabolic turnover of Ac is the rate limiting process, and the Michaelis-Menten kinetics can be used to describe the catalytic process. As observed in Figure S2, S-shape curves were produced, with $E_{KA}$ being measured between near 30 to 50 mV vs. Au. The characteristic S-shape cyclic voltammograms showed a limiting current at high applied potentials between 150 and 250 mV vs. Au. The values of the limiting current increased with age of the biofilm before and during maturity, as expected. As electron transfer in the limiting current portion of the voltammogram indicate efficient electron transfer, 400 mV vs. Au was applied during all chronoamperometry experiments to ensure operation in the current limiting window. In this case the Nernst factor in the Nernst-Michalis-Menten kinetics[8] reduced to 1 (Eq. S7) and manipulation of [Ac] during chronoamperometry should be described by the modified Michaelis-Menten equation (Eq. 4 in the main paper).

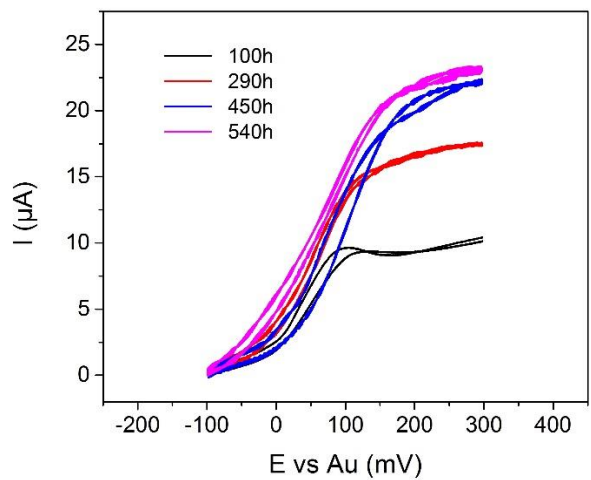

**Figure S2.** CV curves of *G. sulfurreducens* biofilm during growth for 100 h (black curve), 290 h (red curve), 450 h (blue curve) and 540 h (pink curve) after inoculation.

In addition, to the above, we verified the flow independence of the EAB discharge rate as an indicator of flow sensitivity to electron transfer. This is important since flow-dependant conditions, i.e., shear stress, can change by over 10 times in the flow conditions used in this work (See Table S1, section 10). To accomplish this, we looked at the discharge rate for mature biofilms under different flow rates after 10 min exposure to open circuit voltage (OCV) for [Ac] = 0.3 and 10 mM. Immediately following reconnection to chronoamperometic conditions (400 mV vs. Au), current jumped to a maximal value, followed by an exponential decline (Figure S3). The nearly identical discharge curves under different flow rates demonstrated that electron transfer kinetics are not altered during the manipulation of flow and Michaelis–Menten kinetics remain relevant for the description of the rate limiting metabolic turnover of Ac.

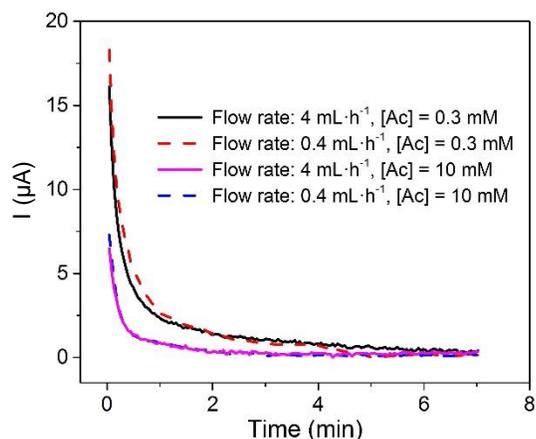

**Figure S3.** Discharge curves on mature biofilm (540 h growth) at 2 different flow rates and nutrient concentrations following 10 min of charging under OCV conditions. Current acquisition was conducted after switching from OCV to 400 mV vs. Au.

## 6. Initial biofilm growth in microfluidic electrochemical cell

Initial electrochemical growth of *G. sulfurreducens* is continuously monitored by CA electrochemical technique from 0 to 140 h (Figure S4). After lag-phase, a rapidly increasing anodic (oxidation) current was measured indicating *G. sulfurreducens* biofilm growth at the WE. The WE was poised on an oxidative potential of 400 mV vs Au (0 V vs. Ag/AgCl). *G. sulfurreducens* biofilm growth in microfluidic electrochemical cell and pseudo-reference electrode studies (stability, electrochemical potential vs. Ag/AgCl and consistency with flow) has been reported before.[2]

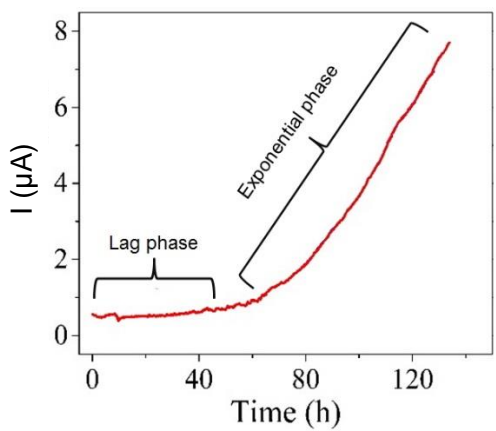

**Figure S4.** Initial electrochemical growth of *G. sulfurreducens*.

## 7. Bulk set-up for bacterial growth and measurements of $K_M$

Figure S5a shows growth curves for *G. sulfurreducens* electroactive biofilm on graphite WE in bulk three-electrode cells. Bulk experiments were done in an electrochemical cell inside an anaerobic chamber. The graphite working electrode (WE) and counter electrode (CE) were cut into 0.5 cm × 0.5 cm for WE and 1 cm × 1 cm for CE. A hole was made by a drill press and a wire was looped and fastened through it and covered by epoxy glue. An Ag/AgCl, 3M KCl was used as a reference electrode. A 100 mL electrochemical cell chamber with 4-hole rubber cap was setup with 3-electrode configuration set-up and a gas tube for purging with 20% $CO_2$ / 80% $N_2$. A 10 mM acetate growth medium, with no fumarate was used for *G. sulfurreducens* growth on the WE at 0 V vs. Ag/AgCl reference electrode. This potential was verified to be the same as 400 mV vs. Au, which was used for microfluidic experiments. After maturation with a steady state current output for 2 days, the nutrient solution was changed with different acetate concentration solutions and the current was recorded for approximately 90 min.

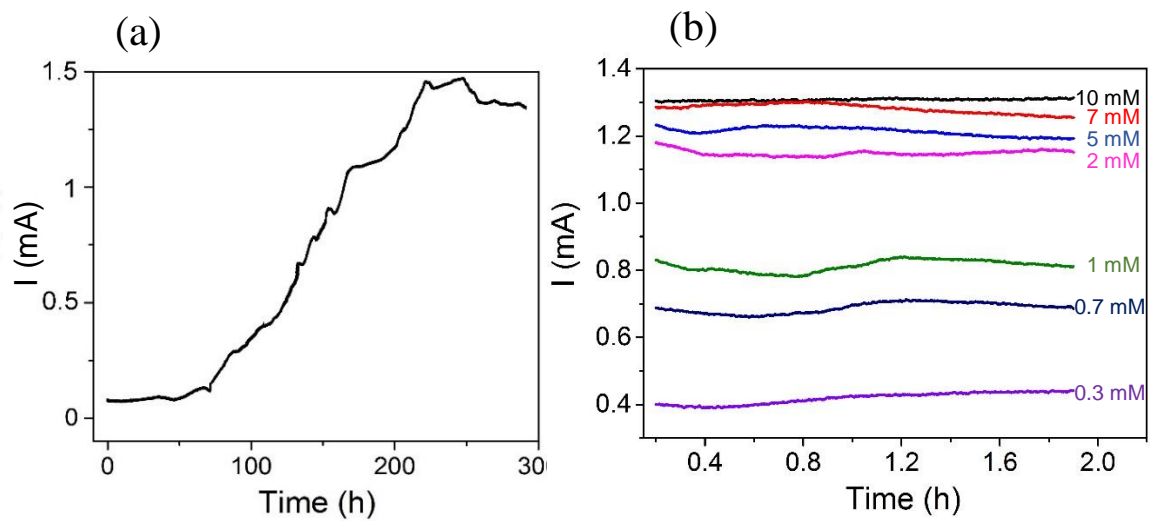

**Figure S5. (a)** CA curve of *G. sulfurreducens* biofilm growth in an electrochemical cell chamber with 10 mM acetate nutrient (WE potential: 0 V vs Ag/AgCl), **(b)** CA curves of *G. sulfurreducens* biofilm at different acetate concentrations. The current output from *G. sulfurreducens* biofilm has been recorded after reaching to steady state conditions for around 90 min.

## 8. Influence of flow on reactor parameters

Table S1 lists several parameters as a function of flow rate, Q. First the average linear channel flow velocity ($\bar{v}_c$) is compared with the average proton diffusion velocity ($\bar{v}_{dH^+}$). This shows that upstream diffusion of protons from the WE where they are generated, is always slower than the average flow velocity. Therefore, the pH at the RE is never affected by down-stream oxidation process, since proton diffusion is faster than any other by-products. The reader is directed to other works using a similar microfluidic electrochemical flow cell, where this point is discussed in more detail.[1,2] Second, calculation of the Reynold's number (Re) shows that flow is strictly laminar at all flow rates in this work. Third, the average shear stress ($\bar{\tau}$) quantifies tangential forces against the electrode-adhered *G. sulfurreducins* biofilm. Finally, the apparent Michaelis-Menten constant, $K_{M(app)}$, and the device reaction capacity, C, are given. These are the result of linear fitting of curve in Figure 3b (main paper), resulting from the Lilly-Hornby equation (Eq. 2, main paper). Calculations for $K_{M(app)}$ and C are derived from the peak I values (Figure 2, main paper), not from the baseline flow rates (Q = 0.2 mL/h).

**Table S1**. Tabulation of changes to mean proton diffusion velocity ($\bar{v}_{dH^+}$), channel flow velocity ($\bar{v}_c$), Reynolds number (Re), shear stress ($\bar{\tau}$), Apparent Michaelis-Menten constant ($K_{m(app)}$) and device reaction capacity (C).

| Q (mL·h$^{-1}$) | $\bar{v}_{dH^+}$ (constant) (mm·h$^{-1}$) | $\bar{v}_c$ (mm·h$^{-1}$) | Re | $\bar{\tau}$ (mPa) | $K_{m(app)}$ (µM) | C (nmol/h) |
|---|---|---|---|---|---|---|
| 0.2 | 8 | 250 | 0.028 | 1.04 | N/A | N/A |
| 0.4 |  | 500 | 0.056 | 2.08 | 572 | 116 |
| 0.6 |  | 750 | 0.083 | 3.13 | 561 | 120 |
| 0.8 |  | 1000 | 0.111 | 4.17 | 551 | 123 |
| 2 |  | 2500 | 0.278 | 10.42 | 521 | 136 |
| 3 |  | 3750 | 0.417 | 15.63 | 466 | 138 |

## 9. Trends in the device reaction capacity

The device reaction capacity, C (mol/h) relates the catalytic rate constant ($k_{cat}$), biocatalytical concentration ([E]) and void fraction ratio β, via $C = k_{cat} \cdot [E] \cdot \beta$. As explained in the main paper, assuming invariability of $k_{cat}$ and β, flow-induced changes to C can be interpreted as changes to [E], the amount of involved bacterial catalyst in contact with the nutrient solution. Figure S6 plots values of C from Table S1 against Q. Non-linear increases to C with Q demonstrate increases to [E]. It is hypothesized that this is due to forced convection through the biofilm. Calculating the $\Delta[C]/[C] \times 100\%$ at Q=0.4 to Q=3, an increase of 22% is noted. Extrapolation backwards to Q=0 is not obvious due to the non-linear nature of C, but an estimation of 109 nmol/h is obtained using the three data points from the lowest flow rates, in the linear portion of the figure. Thus C=138 at 3 mL/h marks an increase of approximately 27% over estimated Q=0 values, and an equivalent increase in contacted *G. sulfurreducens* by the nutrient solution. With the assumption that changes in C are the result of changes in [E] we estimate that elevated flow rates of Q=3 mL·h$^{-1}$ increased [E] by 22% and 27% over Q = 0.4 and static conditions, respectively.

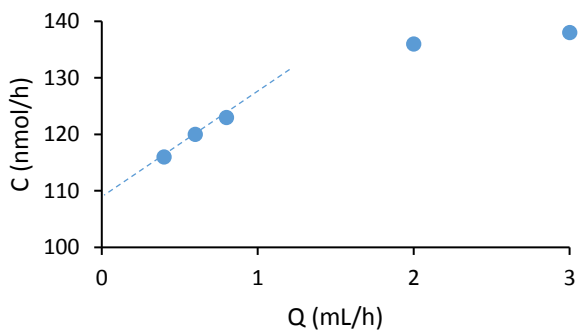

**Figure S6.** Changes in device reaction capacity with flow rate, using data from Table S1. The blue dotted line extrapolates backwards the trends from the linear portion of the curve to static flow conditions at Q=0.

## 10. Trends in $K_{M(app)}$ and reaction mechanisms

The reaction mechanism for this work can be clarified based on the Michaelis-Menten kinetic model. For example, considering Eq. 1 in the main paper, flow-induced reductions to $K_{M(app)}$ indicate a progressive increase in the complex ES with increasing Q. In other words, higher local concentrations of Ac availability at the biofilm/liquid interface can occur due to increased convective flux and related reductions in diffusion barriers, which can result in more efficient Ac capture by the bacteria and rapid complexation with cell-bound enzymes, followed by a final oxidation to electrons and other by-products. Thus, the reason for the increased peak current values in Figure 2 is understood to be the result of higher catalytic reaction rate in Eq. 1, i.e., $d[P]/dt = k_{cat}[ES]$. The use of the Lilly-Hornby model demonstrates both similar and dissimilar trends from other literature examples using flow channels packed with microbead-immobilized enzymes.[3-6] In assaying the effect of flow rate on $K_{M(app)}$ for enzyme systems, less emphasis has been placed on the implications of the underlying reaction mechanisms. For example, other studies have shown that $K_{M(app)}$ becomes reduced with increasing flow rate.[4] This could be because under diffusion-limited conditions flow-induced reductions to diffusion barriers can result in increased concentration near the enzyme and increase the associated enhancement of substrate-enzyme complexation. For those studies in which $K_{M(app)}$ increases with Q, it is probably because reduced liquid/biofilm contact times and the observed accumulation of deposited substrate at the support surface, resulted in less formation ES.[5] We suggest that this observation might be due to slow preliminary complexation (low $k_a$) or fast decomplexation (high $k'_a$). In other studies, $K_{M(app)}$ was observed to remain unchanged with flow,[6] indicating that substrate binding was not rate-limiting.